\documentclass[aps,twocolumn,prb]{revtex4}
\usepackage{graphicx}
\usepackage{amsmath}
\usepackage{amsfonts}
\begin{document}

\title{Crossover between Mean-Field and Ising Critical Behavior
in a Lattice-Gas Reaction-Diffusion Model}

\author{Da-Jiang Liu}
\affiliation{Ames Laboratory (USDOE), Iowa State University, Ames, Iowa 50011;
  e-mail: dajiang@fi.ameslab.gov}
\author{N. Pavlenko}
\affiliation{Institut f{\"u}r Physikalische Chemie und Elektrochemie,
  Universit{\"a}t Hannover, Callinstr. 3-3a, D-30167, Hannover,
  Germany}
\author{J. W. Evans}
\affiliation{Ames Laboratory (USDOE) and Department of Mathematics,
Iowa State University, Ames, Iowa 50011}

\begin{abstract}

Lattice-gas models for CO oxidation can exhibit a discontinuous
nonequilibrium transition between reactive and inactive states, which
disappears above a critical CO-desorption rate.  Using
finite-size-scaling analysis, we demonstrate a crossover from
Ising to mean-field behavior at the critical point, with increasing
surface mobility of adsorbed CO or with decreasing system size.  This
behavior is elucidated by analogy with that of equilibrium Ising-type
systems with long-range interactions.

\end{abstract}

\date{\today}

\maketitle

\section{Introduction}

Nonequilibrium phase transitions and pattern formation occur in a
broad variety of physical, chemical, biological, and sociological
systems.\cite{haken75,mikhailov90} Such complex behavior sometimes
resembles more familiar phenomena in equilibrium systems. However,
there are also some fundamental differences due in part to the lack of
free energy minimization or detailed-balance constraints for
nonequilibrium systems.  The latter feature also provides a challenge
for development of a rigorous theoretical framework.  Nonetheless, the
spatial-temporal behavior of nonequilibrium transitions can be
effectively studied using one of two paradigms.  The first is through
a set of partial differential equations (e.g., reaction-diffusion
equation for chemical reactions), where phase transitions are related
to bifurcations of the underlying nonlinear
dynamics.\cite{haken75,mikhailov90} By default, criticality is
governed by mean-field (MF) behavior, although Ginzburg criteria can
be developed to characterize noise-dominated regimes.\cite{nitzan80}
The second approach utilizes microscopic interacting particle systems
of which nonequilibrium lattice-gas models form an important subclass.
For these models, one can successfully apply techniques such as
finite-size-scaling, and concepts including that of universality
classes familiar from critical phenomena in equilibrium phase
transitions.\cite{marro99}

Many types of nonequilibrium phase transitions have been studied:
continuous transitions to absorbing states~\cite{marro99} (which are
often in the universality class of directed percolation),
discontinuous transitions between reactive and inactive
states,\cite{ziff86,evans02} and order-disorder
transitions,\cite{liu00} all of which occur in adsorption-desorption
or reaction models; driven lattice-gases~\cite{marro99,schmittmann98}
(where the driving breaks detailed balance); dynamic Ising models with
both spin flip and spin exchange dynamics, but at two different
temperatures leading to interesting crossover
phenomena.\cite{gonzalez-miranda87} For some of these systems, it is
possible to derive rigorous reaction-diffusion equations in the
hydrodynamic limit of rapid exchange or
diffusion.\cite{masi85,evans02}

In this paper, we consider specifically surface reaction models, which
incorporate adsorption-desorption and reaction steps, as well as
surface mobility.  Our motivation is that these models not only
contain the essential ingredients to elucidate complex phenomena
observed in real surface catalysis, but also exhibit many fascinating
features of the above idealized models.  On the one hand, they exhibit
complex dynamics and phase transitions associated with the reaction
mechanism (and perhaps with adspecies interactions), but also surface
diffusion rates provide a key parameter which can tune behavior.  In
fact, while many theoretical studies have considered the regime of
limited or zero mobility, it is the hydrodynamic regime of rapid
mobility that is most often realized in experiments.\cite{evans02}

Our focus here is on first-order phase transitions and associated
critical phenomena in models for CO oxidation on metal surfaces,
behavior analogous to that of the ferromagnetic Ising model. The
central question is ``how does critical behavior depend on surface
mobility'', anticipating that MF behavior could well apply in the
hydrodynamic regime, but that a crossover may occur to non-MF behavior
for lower mobility. Through precise finite-size-scaling (FSS) studies
of simulation data, we confirm this prediction and quantify crossover
behavior. We also comment on experimental realization of crossover for
CO oxidation in high-pressure nanoscale systems.

\section{Reaction model: specification and behavior}

We now describe our lattice-gas reaction-diffusion model, where gas
(ads) denote gas-phase (adsorbed-phase) species.  The following steps
are implemented using a square lattice of adsorption sites with
periodic boundary conditions.  (i) CO(gas) adsorbs on single empty
sites at rate $p=p_\mathrm{CO}$ (chosen between 0 and 1) and desorbs
with rate $d$.  (ii) O$_2$(gas) adsorbs dissociatively onto a pair of
diagonally adjacent empty sites at rate $p_\mathrm{O_2}= 1-
p_\mathrm{CO}$, provided all six neighbors are free of O(ads).  This
``eight-site'' rule reflects strong nearest-neighbor (NN)
O(ads)-O(ads) repulsions.  Since O(ads) is treated as immobile, this
adsorption rule ensures that O(ads) never occupy adjacent sites.  (iii)
adjacent CO(ads) and O(ads) react at rate $k$ (set to unity here).
(iv) CO(ads) hops to adjacent empty sites at rate $h$.  This model has
also been discussed elsewhere, e.g., Refs.~\onlinecite{evans02,liu00}.

Conventional kinetic Monte Carlo (KMC) simulations are used to assess
model behavior for finite $h$.  Noting that mobility of CO(ads) is
often very high under ultra-high vacuum conditions, we also perform a
direct analysis of limiting behavior for $h=\infty$ using a ``hybrid''
treatment: here the distribution of O(ads) is described within a
lattice-gas framework, but one only tracks the number of CO(ads) and
assumes that they are randomly distributed on sites not occupied by
O(ads).\cite{evans02} For very large $h$ and low O-coverages, this is
valid.

We now briefly review the steady state behavior of this model for an
infinite system.  \emph{First}, consider the case of finite $h <
\infty$.\cite{ziff86,evans02} For low $d$, one typically finds a
first-order transition at some $p=p_\mathrm{CO}=p^*$ between a
reactive state with low CO-coverage $\langle \theta_\mathrm{CO}
\rangle$ (for $p<p^*$) and an inactive state with high $\langle
\theta_\mathrm{CO} \rangle$ (for $p>p^*$).  The transition disappears
as $d$ increases above some critical value $d_{c}(h)$.  \emph{Second},
consider the hybrid model with $h=\infty$.\cite{evans02,liu00} Here
one finds a region of bistability with both reactive and inactive
states (the discontinuous transition at $p=p^*$ for finite $h$
corresponding to the equistability point). Bistability disappears at a
cusp bifurcation upon increasing $d$ above some $d_c(\infty)$.  A
coherent picture for both cases of infinite and finite $h$ comes from
the observation that decreasing $h$ decreases the degree of
metastability or hysteresis in the system.

Next, we describe how behavior changes for finite $L$.  For the
reaction model with finite $h$, the discontinuous transition in the
$\langle \theta_\mathrm{CO} \rangle$ versus $p_\mathrm{CO}$ mentioned
above is rounded for finite $L$, but becomes sharper as $L$ increases.
The trend is analogous to behavior for the equilibrium Ising model in
finite systems.  In the hybrid model with $h = \infty$, there is also
a smooth transition in the $\langle \theta_\mathrm{CO} \rangle$ versus
$p_\mathrm{CO}$ for finite $L$.  This reflects the feature that for $L
< \infty$, the system can make noise-induced jumps between the low
$\langle \theta_\mathrm{CO} \rangle$ reactive and high $\langle
\theta_\mathrm{CO} \rangle$ inactive states, and that the relative
weight of these states changes smoothly with $p_\mathrm{CO}$.  This
transition also becomes sharper with increasing $L$, occurring at the
equistability point for $L \to \infty$.  See Ref.~\onlinecite{liu02a}
for details.

Finally, to place our study in a broader context, we note that our CO
oxidation model corresponds to a modified version of the
Ziff-Gulari-Barshad (ZGB) model,\cite{ziff86} and its extension with
desorption of CO(ads).  The key modifications in our model are
inclusion of: a) hopping of CO(ads); b) NN exclusion of O(ads); and c)
finite rather than infinite reaction rate.  All these features are
important for realistic modeling of CO oxidation.  However, it is
primarily the \emph{first} feature [hopping of CO(ads)] which impacts
the critical behavior of the reactive-inactive transition studied in
this paper.  Furthermore, it is clear that our conclusions about
critical behavior would apply to the ZGB model, modified to include
desorption and hopping of CO(ads).  The \emph{second} feature [NN
exclusion of O(ads)] results in an oxygen poisoning transition in the
ZBG model\cite{ziff86} being replaced by an order-disorder
transition.\cite{liu00} This does not affect critical behavior of the
reactive-inactive transition, with one caveat.  For small $h$, NN
exclusion does also lead to a loss of the reactive-inactive transition
(see Sec.~\ref{sec:crossover}).  The \emph{third} modification to the
original ZGB model is that instead of $k=\infty$, we choose $k=1$,
i.e., the reaction rate equals the total adsorption rate.  The choice
is of course somewhat arbitrary, but the basic behavior of these
models does not change varying the reaction rate from $O(1)$ to
$\infty$.  For further discussion of effects of varying $k$ on the
steady state bifurcation diagram, see Ref. \onlinecite{evans99}.

\section{Critical point determination and FSS analysis}

Unlike the Ising model, transitions in reaction model do not involve
simple symmetry-breaking.  As in the liquid-vapor phase separation
problem, one needs to locate the critical point in a two parameter
space, i.e., $(p, d)$ for the reaction model.  However here one also
has the disadvantage that computationally efficient techniques for
equilibrium systems (e.g., histogram-reweighting and cluster
algorithms) do not apply.  Thus, when using numerical techniques such
as FSS, careful analysis of simulation data is necessary.  Below, we
briefly describe our procedure.

There are several reasonable ways to define the effective transition
point $p=p^*_L$ for finite $L$.  One natural choice is the point where
the change in the relevant order parameter (i.e., the CO coverage
$\langle \theta_\mathrm{CO} \rangle$) is greatest.  All choices should
converge to the same value as $L$ diverges (for fixed $d<d_c$), and
should yield the same critical exponents in the FSS analysis described
below.  For a given $d<d_c$, we measure CO-coverage $\langle
\theta_\mathrm{CO} \rangle$ versus $p$ for two different system sizes,
say $L$ and $2L$. We find that a convenient definition of $p^*_L$ is
the point where the curves cross, i.e., where $\langle
\theta_{\mathrm{CO}}(p^*_L) \rangle_L = \langle
\theta_{\mathrm{CO}}(p^*_L) \rangle_{2L}$.  The technique is similar
in its underlying motivation to a method for equilibrium first-order
transitions by Borgs and Janke.\cite{borgs92}

After determining the transition pressure $p^*$ for each $d$ value, one
can study the critical behavior using the usual FSS techniques.  For
example, the quantity $\chi_L \equiv L^2( \langle
\theta_{\mathrm{CO}}^2 \rangle - \langle \theta_{\mathrm{CO}}
\rangle^2)$, which is related to susceptibility in equilibrium
systems, is assumed to have the following behavior
\begin{equation}
  \chi_L = L^{\gamma/\nu}
  \tilde{\chi}_\mathrm{Ising}\left[(d-d_c)L^{1/\nu}\right],
\label{eq:ising}
\end{equation}
if the transition belongs to the Ising universality class.  Here
$\nu=1$ and $\gamma=7/4$ are the critical exponents for the
correlation length and susceptibility respectively (in 2D).  In
contrast, for the MF universality class, one has~\cite{mon93,luijten96a}
\begin{equation}
\chi_L = L^{D/2} \tilde{\chi}_\mathrm{MF} \left[(d-d_c) L^{D/2} \right],
\label{eq:mf}
\end{equation}
where $D=2$ is the spatial dimension of the system so that $L^D$ is
the volume (or area in our case) of the system.  Defining $R_L \equiv
\chi_{2L}/\chi_L$, in either case, one has
\begin{equation}
R_L (d_c) \to 2^{\tilde{\gamma}}, \text{ as } L \to \infty
\end{equation}
where the size scaling exponent $\tilde{\gamma}$ is $D/2=1$ for the MF
universality class, and $\gamma/\nu = 7/4$ for the Ising universality
class.

Since one does not know the value of $d_c$ \textit{a priori},
a convenient way to determine both the critical point and exponent
is by finding the crossing point of $R_L(d)$ and $R_{2L}(d)$, so that
\begin{equation}
R_L \left(d_c^L \right) = R_{2L} \left( d_c^L \right) =
2^{\tilde\gamma_L}.
\label{eq:gamma_l}
\end{equation}
$\tilde{\gamma}_L$ can be considered as the effective critical
exponent for finite systems (see Sec.~\ref{sec:crossover}).  Note that
in determining $d_c^L$ and $\tilde{\gamma}_L$, we need simulations of
system of linear sizes $L$, $2L$, and $4L$.

\begin{figure}[!tb]
\centerline{\includegraphics[width=3.25in]{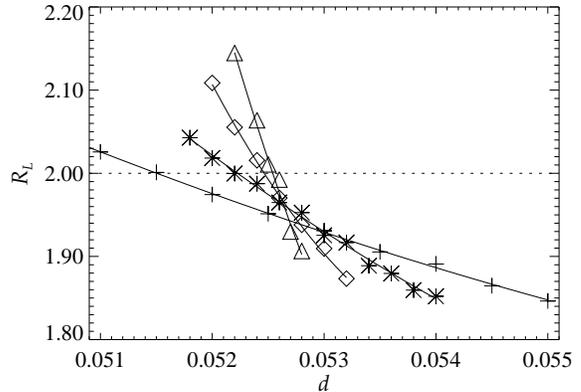}}
\caption{Ratio, $R_L$, of mean-square fluctuation amplitudes for
  systems of size $2L$ and $L$ at equistability points for the hybrid
  model.  $L$ ranges from 8, 16, 32 to 64 with increasing steepness.
  The dotted line plot $R_L = 2$, the MF prediction.}
\label{fig:ri_da2a}
\end{figure}

Figure \ref{fig:ri_da2a} shows $R_L$ versus $d$ for the hybrid model.
The FSS argument above predicts that $R_L$ for different $L$'s will
cross at $(d_c, R_c)$ where $R_c=2$ for MF criticality, and
$R_c=2^{7/4} \approx 3.364$ for Ising (in 2D) criticality.  Figure
\ref{fig:ri_da2a} clearly shows MF behavior for system up to $L=128$.
Note that in the limit of $L \to \infty$, $R_L$ is a step function
with $R_L =4$ when $d < d_c$, and $R_L=1$ when $d >
d_c$.\cite{privman83}

Table \ref{tab:hyb} lists the effective critical exponent obtained
using the above procedure for the hybrid model.  All estimates for
$\tilde\gamma_L$ are close to unity, consistent with the prediction of
MF universality class where $\tilde{\gamma}=1$.\cite{mon93}
Extrapolation to $L=\infty$ assuming a $1/L^2$ finite size correction
(consistent with MF criticality) gives $d_c = 0.05258(5)$ and $p_c =
0.41327(5)$.  Assuming a $1/L$ finite size correction for
$\tilde\gamma_L$ gives $\tilde\gamma=0.995(6)$.  It is possible that
corrections to scaling are described by other exponents.  However, our
analysis shows that the range of system sizes is large enough so that
the value of $d_c$ and the consistency with MF
behavior is not dependent on the form of the corrections.  Note that
MF behavior occurs despite the presence of spatial correlations in the
distribution of O(ads) due to limited mobility of O(ads) and
interactions between O(ads).\cite{liu00}

\begin{table}
  \caption{Effective critical point and critical exponent for
    fluctuations of $\theta_\mathrm{CO}$ for the hybrid model obtained
    from finite-size-scaling analysis.}
  \begin{ruledtabular}
    \begin{tabular}{cccc}
    $L$ & $d_c$ & $p_c$ & $\tilde{\gamma}_L$ \\
    \hline
    8 & 0.05304(6) & 0.4138(1) & 0.948(2) \\
    16 & 0.05275(5) & 0.41343(8) & 0.968(4) \\
    32 & 0.05264(3) & 0.41330(6) & 0.981(5) \\
    \end{tabular}
  \end{ruledtabular}
  \label{tab:hyb}
\end{table}

\begin{figure}[!tb]
\centerline{\includegraphics[width=3.25in]{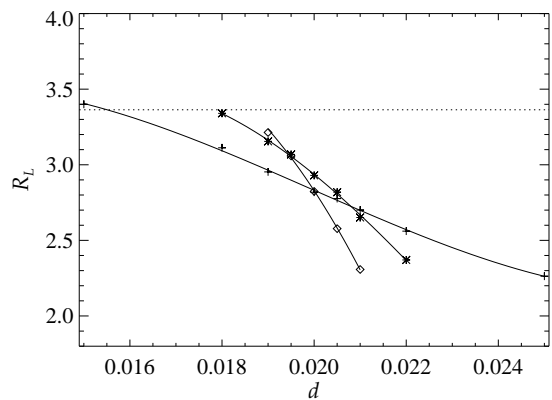}}
\caption{Ratio, $R_L$, of mean-square fluctuations amplitudes for
  systems of size $2L$ and $L$ at equistability points for the
  reaction model with finite CO diffusion model $h=1$.  $L$ ranges
  from 8, 16, and 32 with increasing steepness.  The dotted line shows
  $R_L=2^{7/4}$, the Ising prediction.}
\label{fig:rhdj_da2c}
\end{figure}

Figure \ref{fig:rhdj_da2c} shows $R_L$ versus $d$ for the reaction
model with finite $h=1$. The crossing point appears to approach
$2^{7/4}$ as $L$ increases, consistent with the prediction of Ising
universality.  Limitations in analysis of larger system sizes
precludes definitive convergence of FSS results for either the hybrid
model (Fig.~\ref{fig:ri_da2a} or finite $h$
(Fig.~\ref{fig:rhdj_da2c}.  However, comparing these cases reveals
very different behavior, and we argue that the above assignment of
universality classes is quite reasonable and natural.

From the slope of $R_L$ at the crossing point and the scaling form
$\tilde{\chi}[(d-d_c)L^\theta]$ in Eqs.~(\ref{eq:ising}) and
(\ref{eq:mf}) one can also estimate the exponent related to the
rounding of the critical region due to finite sizes.  Using results in
Fig.~\ref{fig:ri_da2a} we find that $\theta=1.0(1)$ for the largest
system sizes, consistent with the MF value $\theta=D/2=1$.
Uncertainties in the data of Fig.~\ref{fig:rhdj_da2c} are too large to
obtain accurate estimate of $\theta$, but they are consistent with the
Ising value $1/\nu=1$.  In two dimensions, $\theta$ can not be used to
distinguish between the MF and Ising universality class.

We note one previous study by Tom{\'e} and Dickman~\cite{tome93} of
the ZGB lattice-gas model~\cite{ziff86} for CO oxidation (O$_2$
adsorption on adjacent sites, infinitely fast reaction, no CO
diffusion) modified to include CO-desorption.  They measure the shift
of the critical point with system size and found that
$d_c(L)-d_c(\infty) \sim L^{-\lambda}$ where $\lambda=1$ consistent
with results of the two-dimensional Ising model.  Indeed, using a
similar method, we found that $\lambda \approx 1$ for $h = 1$ and
$\lambda \approx 2$ for the hybrid model.\cite{pavlenko03} This result
is consistent with our conclusion regarding universality classes.
However, care must be taken with the interpretation of the shift
exponent, $\lambda$, because it is sensitive to the boundary
conditions (which were chosen to be periodic in both the above
studies) in the case of MF universality where hyperscaling is
violated.\cite{privman83} For example, choosing free boundary
conditions (which modifies adsorption and reaction processes near the
boundaries) could lead to modified $\lambda$ for the hybrid model.

\section{Crossover behavior}
\label{sec:crossover}

Table \ref{tab:rhd_crit} shows the variation with $h$ of the critical
point $d_c$ for the CO poisoning transition obtained using FSS. The
$h=\infty$ value is taken from the hybrid model. By analogy with
equilibrium studies, discussed below and in Ref. \onlinecite{mon93}, we
assume that the shift of $d_c(h)$ away from the limit of $h \to
\infty$ scales as $1/h$ (although there could be logarithmic
corrections).

As an interesting aside, we note that extrapolation behavior to the
regime of small $h$ suggests that $d_c(h)$ vanishes at $h = h_t
\approx 0.2$.  Thus, for small $0 < h < h_t$, the CO poisoning
transition is continuous for $d=0$, and does not exist for $d >
0$. This is in contrast to the ZGB model where a first-order CO
poisoning transition exists even for immobile CO.  For the case of
$d=0$ and $h<h_t$, our epidemic analysis
(cf. Ref. \onlinecite{marro99}) indicates that the transition to the
absorbing (CO poisoned) state, belongs to the directed percolation
universality class.

\begin{table}
\caption{Variation of the critical desorption rate $d_c(h)$ with CO
  hop rate $h$.  Results are obtained from FSS scaling using system
  sizes $L=$16, 32, and 64.}
\begin{ruledtabular}
\begin{tabular}{cccccccc}
$h$ & 0.5 & 1 & 2 & 4 & 10 & 20 & $\infty$ \\
\hline
$d_c(h)$ & 0.012 & 0.019 & 0.027 & 0.035 & 0.042 & 0.047 & 0.0527 \\
\end{tabular}
\end{ruledtabular}
\label{tab:rhd_crit}
\end{table}

Thus, to summarize our results for the critical point in our reaction
model with $k=1$, we find that the critical desorption rate $d_c$
increases from $d_c=0$ for (small) $h=h_t$ to $d_c=0.0527$ for
$h=\infty$. In contrast, for the ZGB model (no NN exclusion of O) with
$k=\infty$, modified to include CO desorption and CO mobility, one
finds that $d_c=0.04$\cite{brosilow92,tome93} for $h=0$ increasing to
$d_c=2/3$ for $h=\infty$.\cite{evans99} Further refining this ZGB-type
model to incorporate $k=1$ (rather than $k=\infty$), one finds that
$d_c=0.03$\cite{evans94a} for $h=0$ increasing to $d_c=0.142$ for
$h=\infty$.\cite{tammaro95}.

Of more central interest to this paper is the variation with $h$ of
the (effective) critical exponents.  Upon increasing the CO diffusion
rate, $h$, how does one crossover from Ising criticality (applying for
finite $h > h_t$) to the MF criticality of the hybrid model (which
corresponds to first taking the limit of $h \to \infty$). A related
question is: for finite (possibly large) $h$, how does critical
behavior depend on finite system size?

In the literature on crossover studies, effective exponents are
usually defined as the slope of the log-log plot of the measured
critical quantities versus deviation from the critical point, e.g.,
$\gamma_\mathrm{eff} \equiv - d \ln \chi / d \ln |d - d_c|$.  A
different definition, which is more amenable to numerical simulations
is to fix the parameters at the critical point, while changing the
system size $L$.  The effective (reduced) exponent can then be defined
as $\tilde{\gamma}_\mathrm{eff} \equiv d \ln \chi / d \ln L$. In
principle, this derivative (or its finite difference approximation)
should be evaluated at the critical point for the infinite system,
$d_c$.  In practice, $d_c$ is unknown \textit{a priori}, and
estimation by extrapolation produces additional errors (particularly
using simulation data for limited system size). Thus instead, one just
uses $d_c^L$ obtained from the FSS analysis in Eq.~(\ref{eq:gamma_l}),
and therefore $\tilde\gamma_L$ also from Eq.~(\ref{eq:gamma_l}) as the
effective exponent.

\begin{figure}[!tb]
\centerline{\includegraphics[width=3.25in]{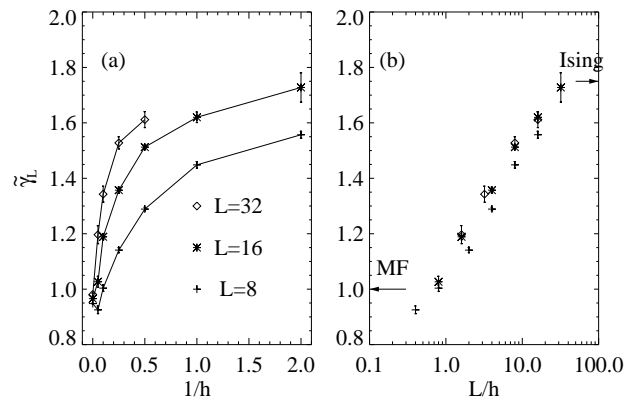}}
\caption{Effective critical exponent $\tilde{\gamma}$ versus
  $h$. See text for details.}
\label{fig:rhd_crit_gamma1}
\end{figure}

Figure~\ref{fig:rhd_crit_gamma1}(a) shows $\tilde{\gamma}_L$ versus
variable $1/h$ for $h=0.5$, 1, 2, 4, and 10.  The system size is
indicated in the figure.  As Fig.~\ref{fig:rhd_crit_gamma1}(a) shows,
$\tilde{\gamma}$ approaches 1 as $h \to \infty$, and approaches 7/4
when $h$ is small. This behavior can be understood in terms of the
theoretical framework presented below.


Phenomenological analysis of nonequilibrium dynamics (and more
specifically of critical phenomena) in Hamiltonian systems with
nonconserved order parameters has traditionally been formulated in
terms of time-dependent Ginzburg-Landau equations
(TDGLE).\cite{hohenberg77} For multi-component order parameter,
$\phi_\alpha$, the TDGLE has the general form
\begin{equation}
\frac{\partial \phi_\alpha(\mathbf{x},t) }{\partial t} = - \Gamma
  \frac{\delta F}{\delta \phi_\alpha(\mathbf{x},t)} +
  \zeta_\alpha(\mathbf{x},t),
\end{equation}
where $\zeta_\alpha$ is Gaussian white noise terms.  Here $F$ is the
coarse-grained free energy functional, which for an Ising-type model
(with single-component order parameter) with interactions of range $R$
can be written as~\cite{mon93,luijten96a}
\begin{equation}
F = \int d \mathbf{x} \left[ V(\phi) + \frac{1}{2} \left|R \nabla
  \phi \right|^2 \right].
\end{equation}
The quartic potential $V(\phi)$ converts from double- to single-well
form with increasing temperature through the critical point.  This
type of formulation extends to nonequilibrium reaction systems
characterized by a vector $\mathbf{c}$ of adspecies concentrations
where the TDGLE are replaced by reaction-diffusion equations (RDE) of
the form
\begin{equation}
\partial \mathbf{c}(\mathbf{x},t) /\partial t \propto - \partial
V_\mathrm{eff}(\mathbf{c}) / \partial \mathbf{c} + \nabla \cdot
\mathsf{D} \cdot \nabla \mathbf{c} + \zeta.
\end{equation}
Analogous to $V$, the effective potential $V_\mathrm{eff}$ converts
from double- to single-well passing through the critical point via the
bistable region.  In our problem, the diffusion tensor satisfies
$\mathsf{D} \propto h$ (the microscopic hop rate) except for small
$h$.\cite{crit_note3} Thus, comparing the RDE with TDGLE after
performing the functional differentiation, it is clear that $h^{1/2}$
plays the role of $R$.  This result is reasonable if one recognizes
these types of reaction-diffusion models (with finite reaction rate
and typically large diffusion rates) exhibit a characteristic
diffusion length, $L_\mathrm{diff}$, which scales like
$L_\mathrm{diff} \sim (h/K)^{1/2}$.  Here, $h$ is the microscopic hop
rate, and $K$ is some effective rate for the overall
adsorption-reaction process.  See
Refs. \onlinecite{evans02,imbihl95}. Thus, refining the above
statement, we can say that $L_\mathrm{diff}$ plays the role of $R$.
Finally, we also note that from analysis of either the TDGLE or the
RDE for infinite systems, it is long known that fluctuations dominate
close enough to the critical point for spatial dimension $D<4$.

Recently, extensive studies have been performed on finite size effects
in equilibrium Ising-type systems with long-range interactions.  From
scaling and renormalization procedures,\cite{mon93,luijten96a} as well
as numerical simulations,\cite{luijten97b} it has been shown that the
crossover from MF to Ising behavior in finite $L \times L$ systems is
governed by the ratio $L/R^2$.  Thus, by analogy, one expects that the
crossover for reaction-diffusion systems is governed by $L/h$ (in 2D).
However, equilibrium and reactions-diffusion systems are quite
different in detail: the Hamiltonian formulation, which is central to
analysis of the former, does not exist for the latter; the microscopic
nature of diffusion is distinct from that of long-range interactions.
Thus, it is appropriate to test numerically the above scaling
hypothesis.  In Fig.~\ref{fig:rhd_crit_gamma1}(b), we replot the data
in Fig.~\ref{fig:rhd_crit_gamma1}(a) using scaling variable $L/h$.
Data collapse, while imperfect for $L=8$, is quite good for $L>16$,
supporting our proposal that that $L/h$ is the correct size scaling
variable.

Of course, it would be preferable to set the phenomenological analysis
on a more rigorous footing. In general, the strategy of extending
mean-field-type formulations to include noise via stochastic partial
differential equations can be unreliable. An alternative rigorous
approach is to attempt to map the exact master equations for the model
onto a stochastic field theory.\cite{taeuber02a} This strategy has
been applied for a variety of toy reaction-diffusion models, e.g.,
demonstrating that the introduction of diffusion can modify the number
of adsorbing states and the universality class in pair-contact models
with diffusion.\cite{hinrichsen00} However, these models are quite
different from our reaction-limited monomer-dimer type models, where
the utility of the rigorous approach has yet to be demonstrated.

\section{Conclusions}

Catalytic CO oxidation on noble metal surfaces provides a convenient
(and important) chemical system for which lattice-gas modeling can be
used to rigorously assess criticality in a reaction-diffusion system.
We have systematically studied the role that CO diffusion plays on the
criticality of the CO poisoning transition. We find a crossover from
MF criticality for fast diffusion [which occurs despite spatial
correlations in the distribution of O(ads)], to Ising criticality for
limited diffusion.  The natural variable describing this crossover is
proposed based on phenomenological arguments.  The proposition is
supported by numerical simulations.

Recent experiments on CO oxidation in nanoscale systems can probe
fluctuations and critical behavior.\cite{suchorski99,suchorski01}
Thus, the considerations of this paper, including the study of
crossover as a function of system size, become particularly
relevant. Furthermore, the crossover behavior described in this study
may be realized physically as one makes the transition from
low-pressure to high-pressure catalysis. The latter could produce a
decrease by many orders of magnitude of the diffusion rates (relative
to other rates) that results from the higher surface coverages.
\cite{crit_note2}

\acknowledgments

DJL and JWE were supported by the USDOE-BES through Ames Laboratory
which is operated for the USDOE by Iowa State University under
Contract No. W-7405-Eng-82.  N.P. gratefully acknowledges the support
from the Alexander von Humboldt Foundation.  We also thank
Prof. Ronald Imbihl for useful suggestions.

\bibliography{bib4}
\end{document}